\documentclass[aps,superscriptaddress,
amsmath, showpacs,twocolumn,prl]{revtex4}

\usepackage{graphicx}
\usepackage{amsfonts}
\usepackage{amssymb}
\usepackage{amsbsy}
\usepackage{amsmath}
\usepackage{latexsym}
\usepackage{natbib}
\usepackage{bm}
\usepackage{subfigure}
\usepackage{color}
\usepackage{hyperref}
\usepackage{lscape}

\newcommand*{\p}{\hat{p}}

\def\p {\partial}

\def\be {\begin{equation}}
\def\ee  {\end{equation}}
\def\bea {\begin{eqnarray}}
\def\eea {\end{eqnarray}}
\def\nn {\nonumber}

\renewcommand*{\k}{\hat{k}}

\renewcommand*{\l}{\lambda_{\star}}

\def\k{\mathbf{k}}
\def\x{\mathbf{x}}
\def\y{\mathbf{y}}

\begin{document}

\title{Propagator in polymer quantum field theory}

\author{Golam Mortuza Hossain}
 \affiliation{ Department of Mathematics and Statistics, University
of New Brunswick, Fredericton, NB, Canada E3B 5A3} 
\affiliation{Department of Physical Sciences, Indian Institute of Science
Education and Research - Kolkata, Mohanpur Campus, PO - BCKV Main Office,
Nadia - 741 252, WB, India}

\author{Viqar Husain}
 
\author{Sanjeev S. Seahra}

\affiliation{ Department of Mathematics and Statistics, University
of New Brunswick, Fredericton, NB, Canada E3B 5A3} 

\date{December 17, 2010}

\begin{abstract}
 We study free scalar field theory on flat spacetime using a background independent (polymer) quantization procedure. Specifically we compute  the propagator using a method  that takes   the  energy spectrum and position matrix elements of the harmonic oscillator as inputs.   We obtain  closed form results in the infrared and ultraviolet regimes  that  give Lorentz invariance violating dispersion relations, and  show  suppression of propagation at sufficiently high energy.    
\end{abstract}

\maketitle

\section{Introduction} 

One of the important problems  in fundamental physics is to understand  the high energy behaviour of quantum fields. This question is intimately connected with  the structure of spacetime at short distances, as the background mathematical structure that underlies quantum field theory (QFT), ie.\ a manifold with a  metric, may come into question in this regime.  A part of the problem is that the spacetime metric forms a reference not only for defining the particle concept, but also for the Hilbert space inner product; if the metric is subject to quantum fluctuations, its use in an inner product becomes an issue.

There are many approaches that have been deployed to probe such questions, including string theory, non-commutative geometry, loop quantum gravity and causal sets. Most of these use  some notion of ``background independence,''   a term which  broadly means that either results are independent of the metric used in calculations, or that a metric is not used at all. 

 In this paper we explore a background independent  (``polymer'') quantization method that arose in loop quantum gravity (LQG) \cite{lqgTT} and apply it to scalar field theory.  In this approach  the  Hilbert space used for quantization is different from the one employed in usual quantum theory. This Hilbert space is such that its  inner product does not make use of a spacetime metric, even if one is available, as in QFT on a fixed background.  Rather the inner product comes from an underlying group structure. This may be compared to the inner product for spin degrees of freedom at each point in a statistical mechanics model.    What is especially interesting about this quantization method is that it introduces a length scale in addition to Planck's constant into the quantum theory. 

The method was applied to the oscillator  \cite{polymer} and other aspects  studied subsequently \cite{Halvorson,HLW-hyd, KLZ}. It is useful to study it further by applying it to QFT on  curved spacetime, which is a question of intrinsic interest. Furthermore  it is needed for a complete theory of quantum gravity, where the same quantization method should be  applied to both geometry and matter variables.  There has already been some work in this direction, such as the construction of  Fock-like states \cite{HK-fock, LV-scalarFT}, an application to matter in cosmology \cite{HHS-cosm}, and a derivation of an effective non-local and Lorentz invariance violating wave equation \cite{HHS-wave}. 

In this work we  develop the area of polymer quantum field theory further by computing 
the  propagator of the scalar field and analyzing its implications for high energy physics. 
We employ an intuitive  approach that directly uses the spectrum of the oscillator. This
introduces a new application of polymer quantization directly in momentum space, and 
gives  some surprising features  not noticed in earlier works such as Ref.\ \cite{polymer}.  
Our results provide an alternative to approaches that   use minimal length  arguments \cite{paddy-prop1} and related background independent methods \cite{SFrev-Baez, SFprop-Rovelli} to compute matter and graviton propagators.
 
\section{Polymer quantum mechanics}
As noted above,  the central difference between Schr\"odinger and polymer quantization is the
 choice of Hilbert space.  The Hilbert  space  used  for the polymer case is the space of almost periodic functions \cite{corduneau}.  A  particle wave function is written as the linear combination 
\be \label{wf}
\psi(p) = \sum_{k=1}^{\infty} c_k e^{ipx_k},   
\ee
where the set of points $\{x_k\}$ is a selection (graph) from the
real line.  In the full Hilbert space, all possible graphs are permitted, so the space
in not-separable. The inner product is
\be \langle x| x'\rangle = \lim_{T\to\infty} \frac{1}{2T}\int_{-T}^T
dp\  e^{-ipx}e^{ipx'} = \delta_{x,x'} , 
\label{ip}
\ee
where the right hand side is the 
generalization of the Kronecker delta to an uncountable index set.  Plane waves are
normalizable in this inner product.

 The configuration operator $\hat{x}$ and translation operator $\hat{U}_\lambda:= \widehat{e^{i\lambda p}}$ act
as
\be \hat{x}e^{i p x_k} = i\frac{\p}{\p p} e^{i p x_k}, \ \ \ \
\hat{U}_\lambda e^{ipx_k} = e^{ip(x_k +\lambda)}. 
\label{basic-rep}
\ee
These operator definitions give the basic commutator
$
 \label{SHOCommBracket}
 [\hat{x}, \hat{U}_{\lambda}] = - \lambda\hat{U}_{\lambda},
$
which is the desired representation of the Poisson bracket $\{x, U_\lambda\} = i \lambda U_\lambda$.
 
The momentum operator cannot  be defined on this Hilbert space because the 
translation operator is not weakly continuous in its  parameter due to the inner
product  (\ref{ip}); only finite translations are realized as in (\ref{basic-rep}). This
is the key feature that leads to modification of energy spectra because the
kinetic energy operator must be defined using the translation operator, which
comes with a fixed scale. Thus all classical observables that depend
on the momentum are realized as scale dependent operators in the quantum theory, 
a feature reminiscent of effective field theory, but coming from an entirely different
source. 
   
Perhaps the simplest definition  of the momentum operator  is 
\be 
\label{eq:p-and-psquared}
\hat{p}_{\lambda} = \frac{1}{2i \lambda} (\hat{U}_\lambda -
\hat{U}^{\dagger}_{\lambda} ) . 
 \ee
In $L_2(\mathbb{R})$, the $\lambda\to0$ limit in (\ref{eq:p-and-psquared})
would give the usual momentum and momentum-squared operators $-i\partial_x$
and~$-\partial^2_x$. In the polymer Hilbert space the $\lambda\to0$
limit does not exist, and $\lambda$ is regarded as a fundamental length
scale. 
The Hamiltonian operator that corresponds to the classical Hamiltonian
$H = p^2/2m + {\cal V}(x)$ is then
\be
\hat{H} = \frac{1}{8m\lambda^2} ( 2- \hat{U}_{2\lambda} -
\hat{U}^{\dagger}_{2\lambda} ) + \hat{\cal V} ,
\label{H}
\ee
where the potential ${\cal V}$ is arbitrary but assumed to be regular so that $\hat{\cal V}$ can
be defined by pointwise multiplication, $\bigl(\hat{\cal V} \psi \bigr) (x) =  {\cal V}(x)
\psi(x)$. One expects the polymer dynamics to be well approximated by the
Schr\"odinger dynamics in an appropriate regime, and certain results to
this effect are known \cite{fredenhagen-reszewski,corichi-vuka-zapata,HHS-wave}.
 
\begin{figure}
 \begin{center}
  \includegraphics[width=\columnwidth]{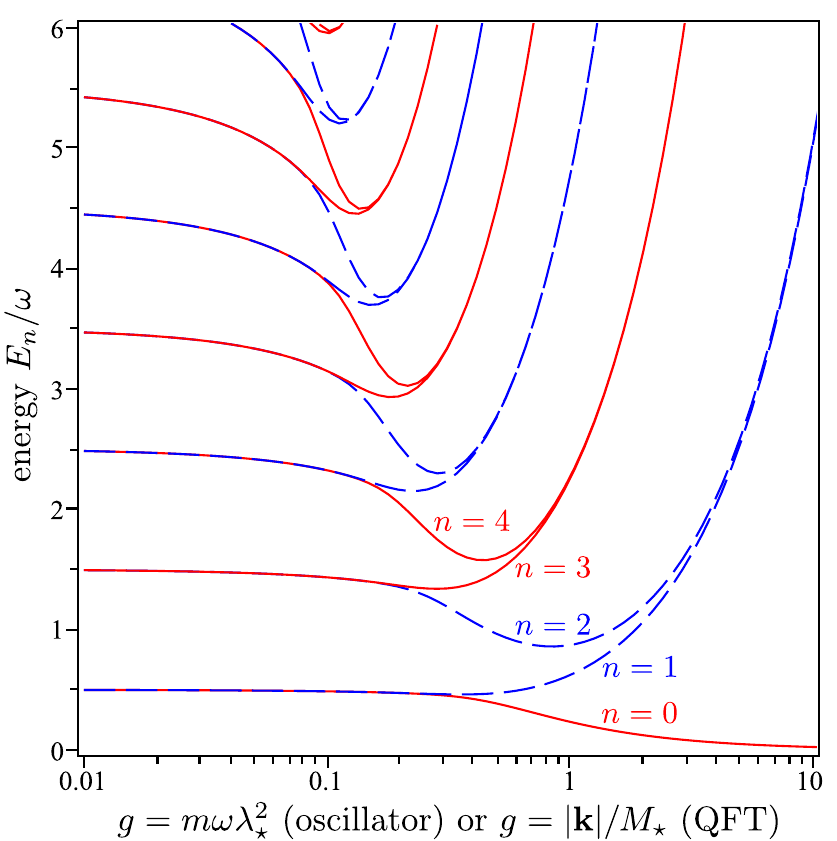}
 \end{center}
 \vspace{-5mm}
 \caption{Energy levels of the polymer oscillator associated with $\pi$-periodic (\emph{red solid}) and $\pi$-antiperiodic (\emph{blue dashed}) eigenfunctions.  The energy levels associated with these two sets are degenerate in the small and large $g$ limits,  with the exception of the ground state energy (labelled $n=0$)  which goes to zero for large $g$. } \vspace{-3mm}
 \label{fig:EnergyLevel}
\end{figure}

We now specialize to the simple harmonic oscillator where ${ \cal V}(x) = m\omega^2x^2/2$.  In the $p$-representation the energy eigenvalue equation $\hat{H}\psi = E\psi$ with the Hamiltonian (\ref{H}) reads: 
\begin{equation}
 \label{PolymerEEEquation}
\frac{1}{8 m\l^2} \left[2 - 2 \cos(2\l p)\right] \psi  -
\frac{1}{2} m \omega^2 \frac{\partial^2\psi}{\partial p^2} = E \psi ~,
\end{equation}
where we have written $\lambda = \lambda_\star$ to denote that  a scale has been fixed. (We use units
where $\hbar=c=1$, so $\lambda$ has dimension of length.)  Defining
\begin{equation}
u = \l p + \pi/2, \quad \alpha=2E/g\omega - 1/2g^2, \quad g= m\omega\l^2,
\end{equation}
transforms (\ref{PolymerEEEquation}) into the Mathieu equation 
\begin{equation}
 \label{MathieuDiffEqn}
  \psi''(u) + \left[\alpha - \tfrac{1}{2} g^{-2} \cos(2u)\right] \psi(u) = 0.
\end{equation}
This has periodic solutions for special values of $\alpha$:
\begin{subequations}\label{eigenfunctions}
\begin{align}
	\psi_{2n}(u) & = \pi^{-1/2}\mathrm{ce}_{n}(1/4g^2,u),         & \alpha & = A_{n}(g), \\
	\psi_{2n+1}(u) &= \pi^{-1/2}\mathrm{se}_{n+1}(1/4g^2,u), & \alpha & = B_{n}(g).
\end{align}
\end{subequations}
Here, $\mathrm{ce}_{n}$ and $\mathrm{se}_{n}$  ($n=0,1\ldots$) are the elliptic cosine and sine functions, respectively, while $A_{n}$ and $B_{n}$ are the Mathieu characteristic value functions \cite{AS-tables}.  For even $n$, $\mathrm{ce}_{n}$ and $\mathrm{se}_{n}$ are $\pi$-periodic, while for odd $n$ they are $\pi$-antiperiodic.  Explicitly, the energy eigenvalues corresponding to the periodic eigenfunctions (\ref{eigenfunctions}) are:
\begin{subequations}
 \label{EigenValueMCFRelation}
 \begin{align}
  \frac{E_{2n}}{\omega} &=  \frac{2g^2 A_n(g)+1}{4g} , &\\
  \frac{E_{2n+1}}{\omega} &=  \frac{2g^2 B_{n+1}(g)+1}{4g}.&
\end{align}
\end{subequations}
We plot these energy levels as a function of $g$ in Fig.\ \ref{fig:EnergyLevel}.  Using asymptotic expansions for $A_n(g)$ and $B_n(g)$, we deduce the following behaviour for $E_n$ in the small $g$ limit: 
\begin{equation}
 \label{EEvalueSmallg}
  \frac{E_{2n}}{\omega} \approx \frac{E_{2n+1}}{\omega} =   \left(n+ \frac{1}{2}\right) -
  \frac{(2n+1)^2 + 1}{16} g + \mathcal{O}\left(g^2\right).
\end{equation}
Clearly,  the Schr\"odinger energy spectrum is recovered for $g=0$ \cite{polymer}. Conversely, in the large $g$ limit we obtain
\bea
\frac{E_0}{\omega} &=& \frac{1}{4g} + \mathcal{O}\left(g^{-3}\right), \\
  \frac{E_{2n-1}}{\omega} &\approx &\frac{E_{2n}}{\omega} =  \frac{n^2 g}{2} + \mathcal{O}\left(g^{-1}\right),
\eea
The first formula shows that the ground state energy falls with increasing $g$, a feature that may have consequences for the cosmological constant problem.  The second formula indicates that the even and odd  energies become degenerate for $g \gg 1$; i.e., when the oscillator mass or frequency is large compared to $\lambda_{\star}^{-1}$.  (The large $g$ behaviour of the system could potentially be used to put experimental bounds on $\lambda_\star$.)  Although different conceptually, it is useful to compare these results to that of an oscillator on a lattice \cite{lattice-osc}.

One of the features of the spectrum of the oscillator is the apparent doubling of energy levels.  
This stems from the fact that the polymer oscillator eigenvalue equation (\ref{MathieuDiffEqn}) is that of a particle in a periodic potential  with period $\pi$. The Hilbert space of solutions is the direct sum ${\cal H} ={\cal H}^+ \oplus {\cal H}^-$, where $\pm$ denotes the $\pi$-periodic and $\pi$-antiperiodic sectors.
One can ask  whether these sectors are superselected, ie. if  $\langle \psi_+ |\hat{A} |\phi_-\rangle =0$ for all operators $\hat{A}$, and for all states $|\psi_+\rangle \in  {\cal H}^+$ and $|\phi_-\rangle \in  {\cal H}^-$. It is readily verified that this is not the case for powers  of the  momentum
operator. 

We note also that  for the alternative kinetic energy operator $(2-\hat{U}_\lambda - \hat{U}_\lambda^\dagger)/\lambda^2$ (used for example in \cite{polymer}), the eigenvalue equation that arises has a periodicity $2\pi$. However this  may be transformed to the standard Mathieu form  (\ref{MathieuDiffEqn})  with the rescaling $\lambda \rightarrow 2\lambda$, and a corresponding scaling of $g$.  The resulting solutions then give rise to the same 2 sectors.  Thus the  doubling of states appears to be a feature of  approach, and independent of the choice of kinetic energy operator. However as we will see below, the $\pi$-periodic sector is naturally selected in the computation of the propagator.

\section{Scalar propagator and dispersion relations}

We turn now to  deriving the propagator of free scalar field theory on Minkowski spacetime using the polymer oscillator spectrum. Defining $g$ as the ratio of the polymer length scale to the wavelength of a given mode [cf.\ (\ref{g-qft})],  we will see that the limit $g\rightarrow 0$ gives the usual Fock quantization results, and the $g\rightarrow \infty$  gives new features which shed  light on the high energy behaviour in polymer QFT.

We start from the standard canonical theory, transform to 3-momentum space $\k$, and apply polymer quantization directly to the Fourier modes of the  phase space variables. This has the known advantage of yielding a Hamiltonian that is a sum of decoupled oscillators. We then compute the two point function directly in the energy eigenfunction basis.  

The phase space variables for the free scalar field are the canonical pair ($\phi(\x,t), \pi(\x,t)$)
satisfying the Poisson bracket
\be
\label{PositionSpacePB}
\{\phi(t,\x), \pi(t,\y)\} = \delta^{(3)}(\x-\y) ~.
\ee
The Hamiltonian is
\be
\label{SFHamGen}
H_{\phi}  =  \int d^3\x \left[ \frac{\pi^2}{2\sqrt{q}} +
\frac{\sqrt{q}}{2} q^{qb} \nabla_a\phi \nabla_b\phi
\right],
\ee
where we have written the metric as $ds^2 = -dt^2 + q_{ab}dx^adx^b$.
We now restrict  to  a flat 3-space, ie. $q_{ab} =e_{ab}$,
with volume $V = \int d^3x \sqrt{q}$.  Thus our field configurations are
those of system in a box.

We transform to 3-momentum space with the Fourier expansion
 \bea
 \label{FourierModesDef}
\phi(t,\x) &=& \frac{1}{\sqrt{V}} \sum_{\k}{\phi}_{\k}(t) e^{i {\k}\cdot{\x}}, \nn\\
\label{mode}
{\phi}_\k(t) &=& \frac1{\sqrt{V}}\int d^3x\ e^{-i\k\cdot x} \phi(\x,t),
\eea 
with a similar expansion for $\pi(\x,t)$.  After  a suitable redefinition of the independent modes  
the  Hamiltonian is 
\be
H_{\phi} =  \sum_{\k} H_{\k} = \sum_{\k} \left[ \frac{\pi_{\k}^2}{2} +
\frac{1}{2} k^2 \phi_{\k}^2 \right],
\ee
with the Poisson bracket $\{\phi_\k, \pi_{\k'}\} = \delta_{\k\k'}$.
The  polymer quantization of each mode now follows that of the oscillator with obvious identifications.  
We use the variables  $\phi_\k$ and $U_{\lambda\k }= e^{i\lambda \pi_k}$ which satisfy the 
same Poisson bracket 
$
\{ \phi_\k, U_{\lambda\k}\} = i \lambda U_{\lambda\k }
$
[note that $\lambda$ now has dimensions of $(\text{length})^{1/2}$].  Taking  the definition of the Hamiltonian operator as in (\ref{H}), and the identifications
$m=1$ and $\omega=|\k|$, the spectrum of $H_\k$ is the same as that  obtained above with 
\be
\label{g-qft}
 g=\lambda_\star^2 |\k|  \equiv \frac{|\k|}{M_\star} \propto \frac{\text{polymer length scale}}{\text{spatial wavelength}},
\ee 
where $M_{\star}^{-1}$ is the fundamental length scale associated with the polymer quantization of $\phi$.

We now show how the polymer oscillator spectrum leads to a modified propagator for $g\ne 0$.
The usual  $2-$point function is 
\be
\label{KGPropagator}
 \langle 0| \hat{\phi}(\x,t)
\hat{\phi}(\x',t')|0\rangle
\equiv  \frac{1}{V} \sum_{\k} e^{i {\k}\cdot(\x-\x')}D_{\k}(t-t') ,
\ee
where $|0\rangle=\Pi_{\k}\otimes |0_\k\rangle$ is the Fock vacuum.  For our case the matrix element in the last
equation is 
\begin{equation}\label{pprop-def}
D_{\k}(t-t') = 
\langle 0_{\k}| e^{i\hat{H}_{\k}t} \hat{\phi}_{\k} e^{-i\hat{H}_{\k}t}
e^{i\hat{H}_{\k}t'} \hat{\phi}_{\k} e^{-i\hat{H}_{\k}t'}
|0_{\k}\rangle,
\end{equation}
where $\hat{H}_\k$ is the polymer Hamiltonian operator
\be
\hat{H}_\k = \frac{1}{8\lambda^2} \left[ 2- \hat{U}_{2\lambda\k } -
\hat{U}^{\dagger}_{2\lambda\k } \right] + \frac{1}{2}\k^2 \phi_\k^2 ,
\label{Hk}
\ee
and $|0_\k\rangle$ is its ground state.  The matrix element in   (\ref{pprop-def}) is readily computed using  the polymer  oscillator spectrum  $\hat{H}_{\k}|n_{\k}\rangle = E_n^{(\k)}|n_{\k}\rangle$,  and the expansion of the state $\hat{\phi}_{\k} |0_{\k}\rangle$ in the energy
eigenstates as $\hat{\phi}_{\k} |0_{\k}\rangle = \sum_{n} c_n |n_{\k}\rangle$, with
\begin{equation}
c_n = \langle n_{\k}| \hat{\phi}_{\k} |0_{\k}\rangle =\lambda_\star \int_0^{2\pi} \psi_n(i\p_u)\psi_0 du.
\end{equation}
Eq.\ (\ref{pprop-def}) then becomes 
\begin{equation}\label{KGPropagator3}
D_{\k}(t-t') = \sum_{n} |c_n|^2 e^{ -i \Delta E_n (t-t')},
\end{equation}
where $\Delta E_n \equiv E_n^{(\k)} - E_0^{(\k)}$. To bring this expression into a more recognizable
form we write the exponential as an integral to give 
\begin{multline}
\label{KGPropagator4}
D_{\k}(t-t') \equiv \int \frac{d \omega}{2\pi} \ D_p\ e^{-i \omega(t-t')} \\
 = \sum_{n} 2 \Delta E_n|c_n|^2  \int\frac{d \omega}{2\pi} 
 \frac{ie^{-i \omega(t-t')} }{\Delta E_n^2 -\omega^2-i\epsilon} ,
\end{multline}
where  $\omega$  here is defined as  the time component of the 4-momentum $p \equiv(\omega,\k)$. 
Thus  the momentum space propagator is  
\begin{equation}\label{KGPropagatorMomentum}
  D_{p} = \sum_{n} \frac{2i\Delta E_n |c_n|^2 }{p^2 + 
  \Delta E_n^2 - |\k|^2 - i\epsilon},
\end{equation}  
where we used $p^2=-\omega^2 +|\k|^2$  and chose the sign of $i\epsilon$ to correspond to the Feynman propagator.  We note that for the Schr\"odinger oscillator  $c_n = \delta_{1,n}/\sqrt{2|\k|}$ and
$\Delta E_n = n |\k|$, so (\ref{KGPropagatorMomentum}) gives the expected result  
$D_{p} = i/(p^2 - i\epsilon)$.

The polymer corrections we seek arise from the dependence of the matrix element $c_n$ and the 
energy differences $\Delta E_n$ on the parameter $g=|\k|/M_\star$. Unlike the Fock case, the sum in  (\ref{KGPropagatorMomentum}) contains an infinite number of terms for $g\ne0$, and appears to be analytically unsummable. However it is possible to obtain formulas for $g\ll 1$ and $g\gg1$ using the appropriate asymptotic representations of the elliptic functions.  We find that the only non-zero matrix elements are $c_{4n+3}$ (for  $n=0, 1, 2, \ldots$).  This is expected because 
the natural ground state $\psi_0$ is even and  $\pi$ periodic, and the operator $\phi_\k$ connects it only to (odd) $\pi$ periodic states. Thus our propagator calculation makes use of only  the $\pi$-periodic eigenfunctions.

\paragraph{The infrared limit $g\ll 1$:}  In this case we find
\be
\Delta E_{4n+3}  = |\k| \left[ (2n+1) - \frac{(4n+3)^2-1}{16} g  + \mathcal{O} \left( g^2 \right) \right] 
\ee
for $n \ge 0$, and 
\bea
\label{SmallgDeltaE4n+3}
  c_3 &=& \frac{-i}{\sqrt{2|\k|}} \left[1 - \frac{3}{4} g
  +\mathcal{O}\left(g^2\right) \right], \\
  \frac{c_{4n+3}}{c_{3}} &=& \mathcal{O}\left(g^n\right),
\eea
for $n >0$.  From these, it is apparent that  only the $c_{3}$ term in (\ref{KGPropagatorMomentum}) gives a contribution to $D_{p}$ that is linear in $g$.   Retaining only this term gives 
\be
 D_p = \frac{i(1-2g) }{p^2   - g|\k|^2 - i\epsilon }  + {\cal O}(g^2).
\ee
Comparing this with the propagator of the massive scalar theory,  we see that (i) the correct $g=0$ limit is obtained, (ii) there exists a tachyonic pole $p^2=g|\k|^2$,  and (iii) this pole implies an effective dispersion relation 
\be
\omega^2 =|\k|^2(1 -|\k|/M_\star)
\ee
 that violates Lorentz symmetry. We note that unlike the case of tachyons in Lorentz invariant theories, this polymer quantization correction does not lead to complex $\omega$ for real $|\k|$, so there is no instability. 
  
 \paragraph{The ultraviolet limit $g\gg1$:} In this limit we find
 \be
 \label{LargegDeltaE4n+3}
  \Delta E_{4n+3} =  |\k| \left[ 2(n+1)^2 g + 
  \mathcal{O}\left(\frac{1}{g^3}\right) \right],
  \ee
  for $n \ge 0$, and 
 \bea
  \label{LargegC4n+3}
  c_{3} &=& i\sqrt{\frac{g}{2|\k|}} \left[\frac{1}{4g^2} +
  \mathcal{O}\left(\frac{1}{g^6}\right) \right], \\
  \frac{c_{4n+3}}{c_{3}} &=& \mathcal{O}\left(\frac{1}{g^{2n}}\right),
\eea
for $n > 0$.  Since the coefficients higher than $c_3$  are suppressed by $1/g^{2n}$ (for $n>1$),  keeping only this term in the propagator sum   (\ref{KGPropagatorMomentum})  gives 
\be
D_p =  \frac{i/8g^2}{p^2 + 4 g^2|\k|^2-i\epsilon} +{\cal O}\left(\frac{1}{g^6}\right).
\ee 
This has the following interesting features: (i) the pole is not tachyonic, unlike the $g\ll 1$ case, (ii) 
the associated dispersion relation 
\be
\omega^2 = 4|\k|^4/M_\star^2 
\ee
 still violates Lorentz invariance  and reflects a higher derivative term, and (iii) the propagation amplitude at high momentum is suppressed by a factor $1/g^2$, a  feature not present in linear higher derivative theory. 

\section{Summary}

We applied the polymer quantization method to the oscillator in momentum space and computed its full spectrum.  We then used this to compute the scalar field propagator using an intuitive approach that exploits directly the decomposition of free field theory into a collection of simple harmonic oscillators. The resulting propagator violates Lorentz  invariance for $g\ne0$. Its limits in the  low and high momentum regime exhibit vastly different behaviour. At low momenta the theory acquires an effective (tachyon) mass $m_{\text{eff}}^2 = -|\k|^3/M_\star$. At large momenta the effective mass is  $m_{\text{eff}}^2= 4|\k|^4/M_\star^2$ and the propagation amplitude is  suppressed. 

Our results provide an example of the type of effects one can expect in  quantum field theory in general curved spacetimes using this quantization method.  In particular it opens up new directions for investigation such as the spectrum of cosmic microwave background fluctuations and  Hawking radiation, where the  scale associated with the quantization is likely to give new physics.

\medskip

\begin{acknowledgements}

{\it Acknowledgements} We thank Gabor Kunstatter and Jorma Louko for helpful comments on the oscillator spectrum.  
This work was supported by the Atlantic Association for Research in the Mathematical Sciences (AARMS) and NSERC of Canada. 

\end{acknowledgements}

\bibliographystyle{apsrev}

\vfill

\end{document}